\documentclass[10pt,letterpaper,twocolumn]{article} 

\usepackage{ol}
\usepackage{amsmath}

\begin{document}

\twocolumn[ 

\title{Optical ratchets with discrete cavity solitons}

\author{Andrey V. Gorbach, Sergey Denisov, and Sergej Flach}

\address{Max-Planck-Institut f\"ur Physik komplexer Systeme, N\"othnitzerstr. 38, Dresden 01187, Germany}


\begin{abstract}
We propose a
setup to observe soliton ratchet effects using 
discrete cavity solitons in a one-dimensional
array of coupled waveguide optical resonators. The net motion of solitons can be generated by 
an adiabatic shaking of the holding beam
with zero average inclination angle. The resulting soliton velocity can be controlled by different 
parameters of the holding beam.
\end{abstract} 


] 

\noindent 
Spatial solitons in optical nonlinear cavities have attracted a lot of attention during last decades, see, e.g.,
reviews in Refs. \citeonline{Rev_Firth, Rev_Pesh} and references therein. 
They represent nondispersive localized structures supported by a flat holding beam (pump) and result 
from the interplay between nonlinearity and diffraction, as well as gain, losses and internal feedback 
of the system.
Due to multiple reflections of light at the boundary mirrors, 
which form the cavity, light interaction with the nonlinear material inside the cavity is effectively increased. 
As a consequence, spatial cavity solitons (CSs) may be formed at essentially reduced input powers, 
as compared to conventional spatial solitons in waveguides.
All the parameters of cavity solitons (including energy and phase) are completely determined
by the parameters of the holding beam, which adds more flexibility to the control of 
the process of creation and annihilation 
of CSs as well as to their evolution.
Together with a high robustness, usually inherent for solitons, 
the above properties turn CSs into good candidates for information storage units in all-optical devices.
Cavity solitons have been experimentally observed in semiconductor microcavities \cite{Exp_Nature} and single-mirror feedback
loops with nonlinear elements \cite{Exp_PRL}. 

\begin{figure}
\centerline{
\includegraphics[width=0.4\textwidth]{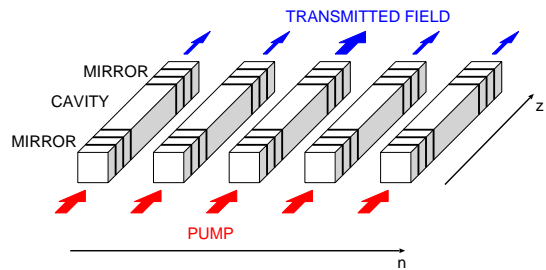}
}
\caption{Schematic setup of coupled waveguides with mirrors at the end facets.}
\label{fig1}
\end{figure}

The presence of gain and losses in the system introduces several qualitatively new features to the 
soliton dynamics.
In particular, a unidirectional soliton motion under the influence of AC 
zero-mean external forces
can be realized, provided certain symmetries of the system are broken \cite{Flach, Salerno}. 
Such a \emph{soliton ratchet} effect has been successfully observed in experiments 
with an annular Josephson junction, revealing 
a unidirectional topological fluxon motion driven by biharmonic microwaves of zero average \cite{Ustinov}. 
Up to now the soliton ratchet
effect has been discussed only for solitons bearing a non-zero topological charge. 
In such a case the excitation can not be destroyed due to large energy barriers
in the system.
In this Letter we extend the 
concept of soliton ratchets 
to much more fragile optical cavity solitons having 
\emph{zero topological charge}.
We consider 
a nonlinear optical waveguide array with dielectric mirrors at the end facets 
(array of coupled zero-dimensional optical resonators), 
driven by an external field (pump), see Fig.~\ref{fig1}.
Recently it was demonstrated that \emph{discrete cavity solitons} (DCSs) 
can be excited in such a system within a reasonably
wide range of control parameters \cite{Pesh_ol}. The  
evolution of the transmitted field pattern at the output facet
obeys a discrete nonlinear Schr\"oedinger (DNLS) type equation \cite{Pesh_ol}:
\begin{eqnarray}
\label{DNLS}
\left(i\frac{\partial}{\partial \tau}+\Delta+i+\alpha |A_n|^2\right)A_n+ \qquad \qquad\\
\nonumber
\qquad \qquad C(A_{n+1}+A_{n-1}-2A_n)=A^{in}_n\;,
\end{eqnarray}
where $A_n$ and $A^{in}_n$ correspond to the renormalized amplitudes of transmitted 
and input fields in the $n$th waveguide, 
respectively. The time $\tau=t/T_0$ is measured in terms of the photon lifetime 
inside the cavity $T_0=2d/[v_g(1-\rho^2)]$,
which may be of the order of picoseconds or larger. Here
$d$ is the length of each waveguide, $v_g$ is the group velocity of light inside the waveguides and $\rho$ is 
the real-valued reflection coefficient of the mirrors, $(1-\rho^2)\ll 1$. The effective damping coefficient  
is rescaled to unity in Eq.~(\ref{DNLS}),
while parameters $C=C_0/(1-\rho^2)$ and $\Delta=\Delta_0/(1-\rho^2)+2C$ define 
the effective coupling between adjacent waveguides
and detuning from the linear resonance, respectively ($\Delta_0=2\beta_0 d$, 
where $\beta_0$ is the wave number corresponding to the laser frequency
$\omega_0$). Both parameters can vary in a wide range by adjusting the reflectivity 
of mirrors $\rho$, as well as the frequency of incident light and the
distance between adjacent waveguides (the latter determines the value of $C_0$). 
The nonlinear Kerr coefficient $\alpha$ can be rescaled to $\alpha=\pm 1$.
Increasing the coupling parameter $C$ the model in Eq.~(\ref{DNLS})
asymptotically approaches the continuous Lugiato-Levefer model \cite{Lugiato} 
for a one-dimensional resonator based on a homogeneous-core slab waveguide. 
Then the discrete Laplacian in Eq.~(\ref{DNLS}) is replaced
by the second spatial derivative $\partial^2/\partial x^2$ of a continuous field $A(x,\tau)$.

\begin{figure}
\centerline{
\includegraphics[angle=270, width=0.35\textwidth]{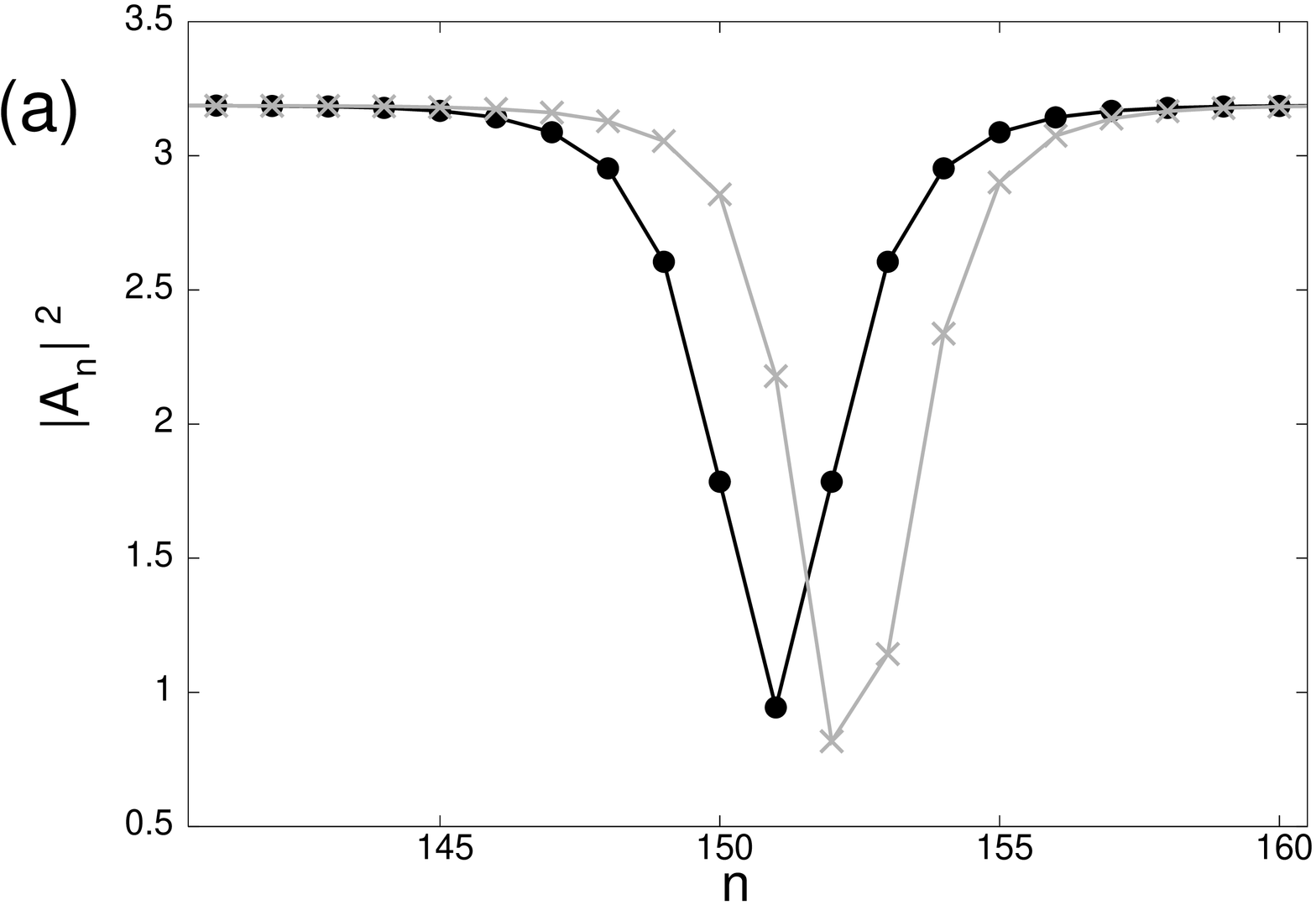}
}
\centerline{
\includegraphics[angle=270, width=0.35\textwidth]{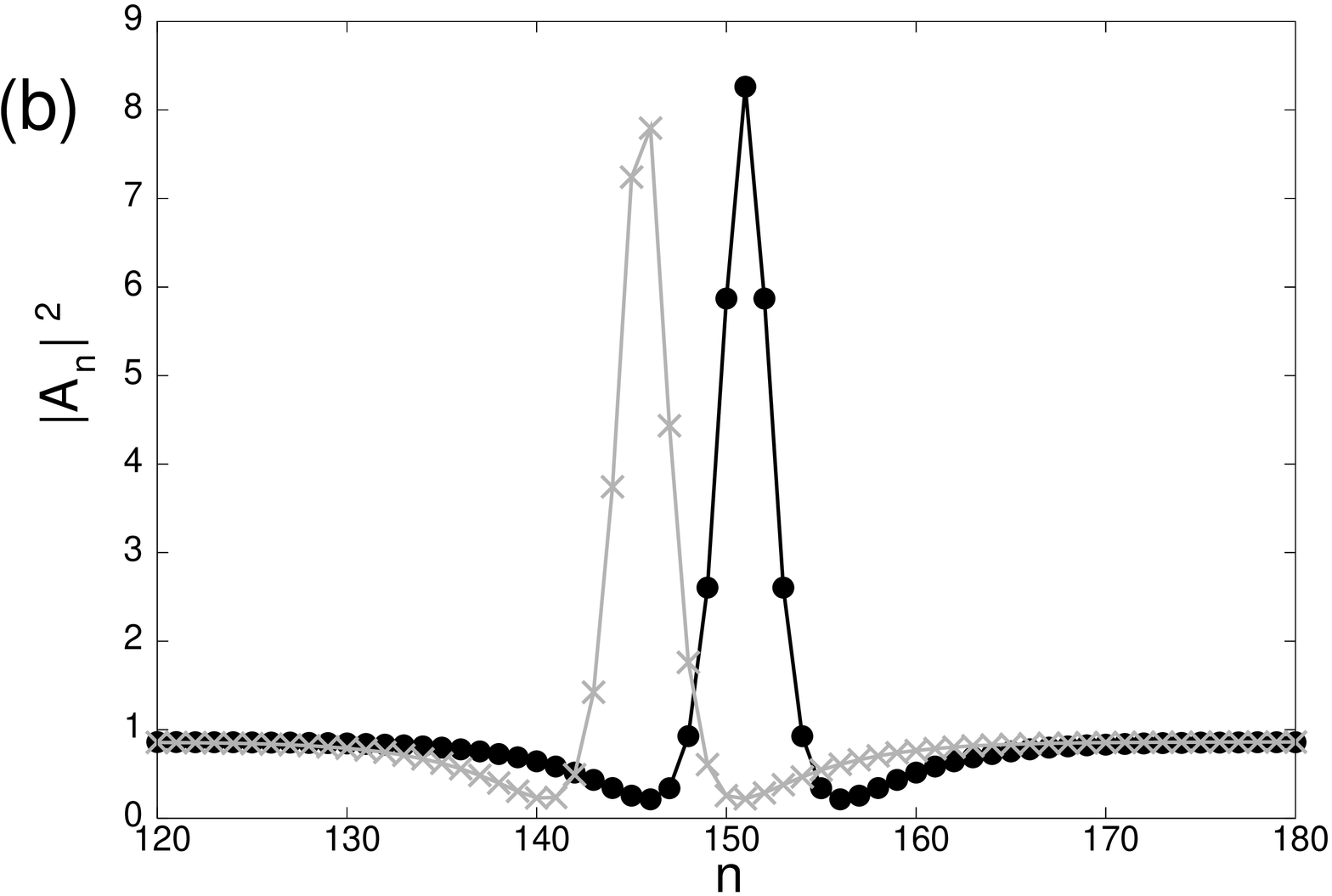}
}
\caption{
Dark (a) and bright (b) DCSs profiles. Black lines and circles:
stationary DCSs for the input power $|a|^2=3.3$ at normal incidence $\phi_{in}=0$.
Model parameters are: (a) $\alpha=-1,\; \Delta=3,\; C=0.55$; 
(b) $\alpha=1,\; \Delta=-3,\; C=15$.
Gray lines and crosses: snapshots of moving DCSs for the ratchet case with the same parameters 
as in Fig.~\ref{fig3}.
}
\label{fig2}
\end{figure}

For a constant input field 
\begin{equation}
A^{in}_n=a\exp(i\phi_{in} n)\;,
\label{const_force}
\end{equation}
$\phi_{in}$ determines the incident angle of the pump light.
Normal incidence $\phi_{in}=0$ Eq.~(\ref{DNLS}) supports stationary DCS solutions \cite{Pesh_ol} 
in a certain regime of the input power $|a|^2$ and detuning $\Delta$, see Fig.~\ref{fig2}. 
%
A tilted holding beam with the incidence angle $\phi_{in}\ne 0$ induces a 
transverse force acting on the soliton.
It results in a moving soliton which can be observed 
both in continuous \cite{Fedorov} 
and discrete \cite{Pesh_mov} models. 

Let us now consider the case, when the incidence angle of the holding beam is adiabatically slowly
oscillating in time with zero mean:  $\phi_{in}(\tau+T)=\phi_{in}(\tau)$ ($T\gg 1$) ,
$\int_\tau^{\tau+T}\phi_{in}(\tau)d\tau=0$.
We want to study whether 
the corresponding induced AC zero mean force can result in a unidirectional 
(in average) ratchet motion of the soliton, 
disregardless the initial conditions (i.e. the initial phase of $\phi_{in}(\tau)$, 
position of the soliton, etc.).
For the period of the holding beam oscillations $T$ being much larger than the characteristic 
photon lifetime in the cavity, $T \gg 1$,
the soliton dynamics is locked to the beam phase, i.e. there exist only a unique attractor solution. 
As a first step 
we perform a symmetry analysis 
\cite{Flach} of Eq.~(\ref{DNLS}).  Suppose we identify a symmetry operation $\hat{S}$, 
which involves shifting of spatial and temporal coordinates
and/or mirror reflections, leaves the dynamical equations invariant and, at the same time, 
changes the sign of the soliton velocity. If applied to any trajectory,
it will again generate a trajectory.
Since the attractor is unique, the symmetry operation applied to the attractor
will map it onto itself, and the average velocity on this attractor
will be exactly zero, so that the
soliton will perform periodic oscillations in space and time.
%
%
Therefore, breaking such symmetries of the underlying model equations 
is a necessary condition to observe a soliton ratchet effect.

\begin{figure}
\centerline{
\includegraphics[angle=270, width=0.4\textwidth]{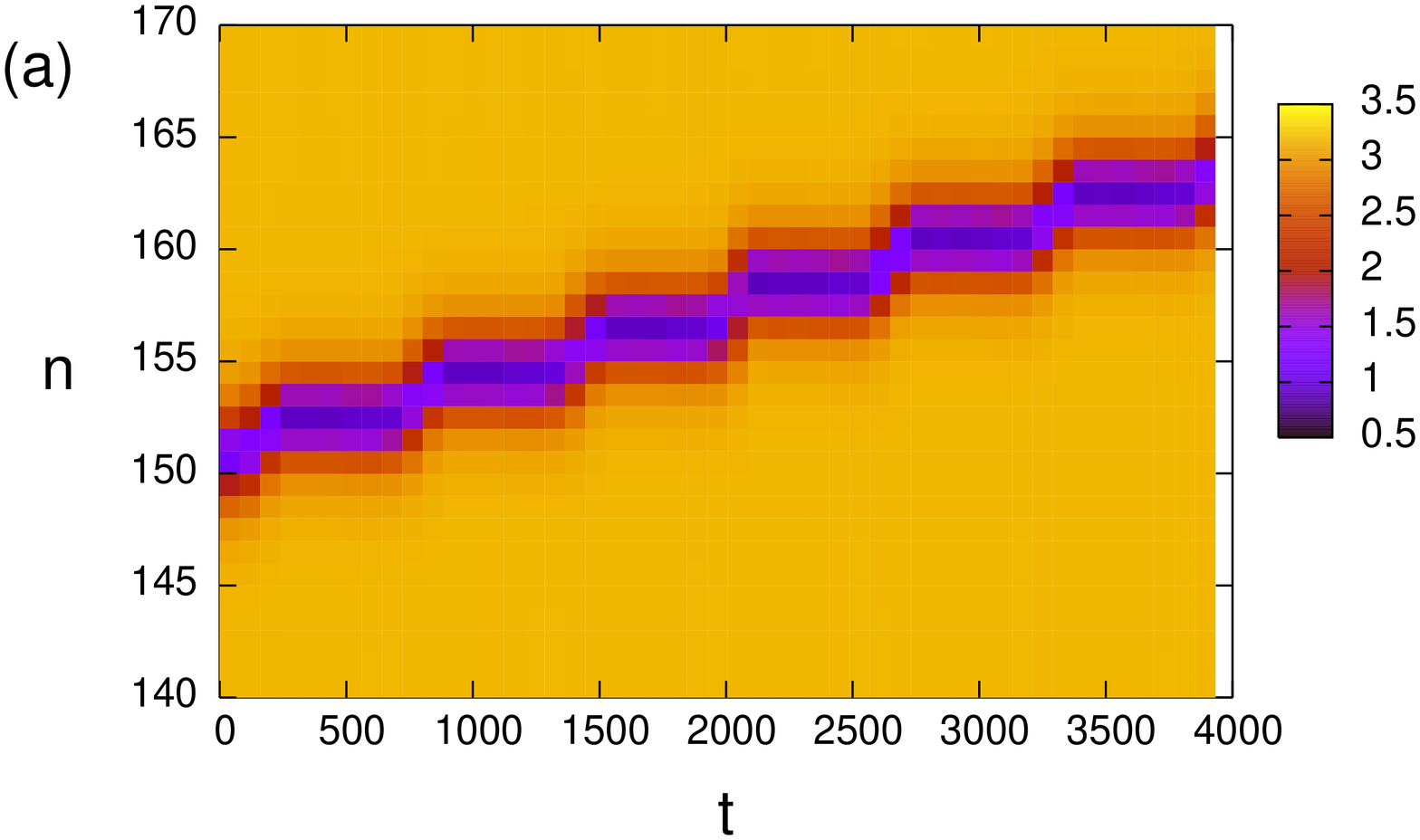}
}
\centerline{
\includegraphics[angle=270, width=0.4\textwidth]{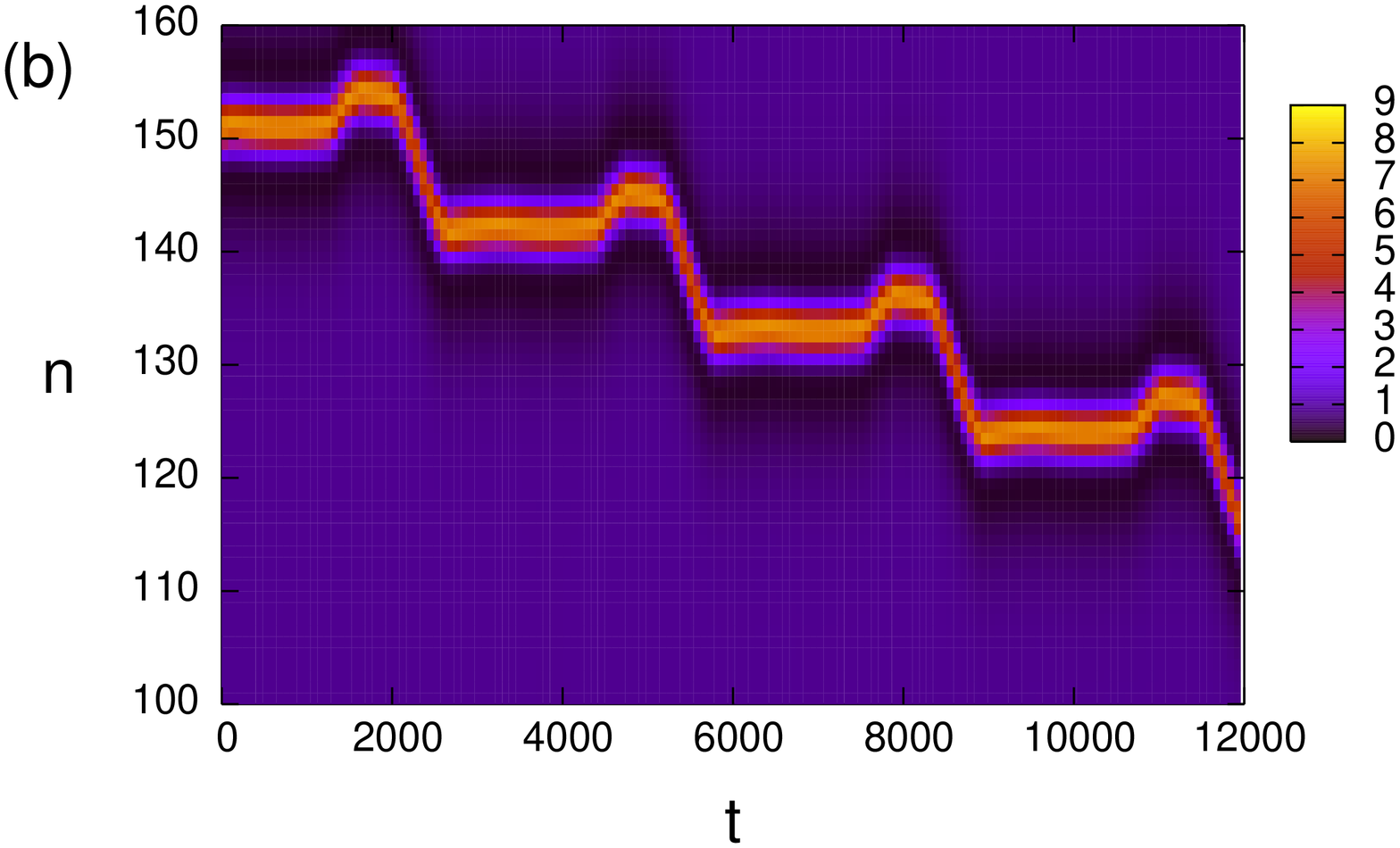}
}
\caption{Density plots of $|A_n|^2$ for soliton motion 
in the ratchet case with initial conditions corresponding to stationary
DCSs. Model parameters are the same as in Fig.~\ref{fig2}.
(a) Dark DCS driven by the single shaking beam~(\ref{const_force}), (\ref{biharm1}), 
$|a|^2=3.3$,
$\phi_a=\phi_b=0.025$, $\omega=0.01$, $\theta=1.72\cdot \pi$.
(b) Bright DCS driven by two shaking beams~(\ref{biharm2}),
$|a|^2=|b|^2=1.2$, 
$\phi_a=\phi_b=0.001$, $\omega=0.002$, $\theta=0.57 \cdot \pi$.
}
\label{fig3}
\end{figure}

\begin{figure}
\centerline{
\includegraphics[width=0.35\textwidth]{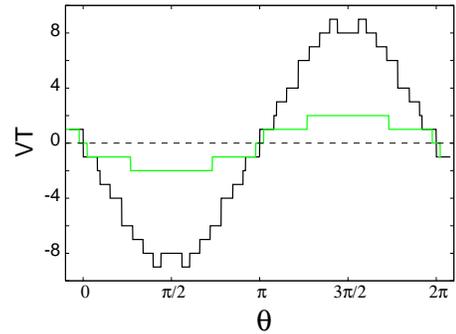}
}
\caption{Soliton displacement per period $T$ as a function of the relative phase $\theta$ for dark (green line) and
bright (black line) solitons. All the parameters are the same as in Fig.~\ref{fig3} (a) and (b), respectively.
}
\label{fig4}
\end{figure}

Far away from the soliton center the field distribution asymptotically 
approaches the homogeneous (in intensity) ground state.
The corresponding ground state field distribution 
$A^{GS}_n(\tau)\approx a^{GS}(\tau)\exp(i\phi^{in}n)$ and its intensity 
$I^{GS}(\tau)=|a^{GS}(\tau)|^2$ are obtained from \cite{Pesh_ol}
\begin{equation}
\left[\Delta-4C\sin^2(\phi_{in})+i+\alpha|a^{GS}|^2\right]a^{GS}=a\;.
\end{equation}
Then, we can define the position of the center of soliton $X(\tau)$ and its velocity $V(\tau)$
\begin{equation}
\label{center}
X(\tau)=\frac{\sum_n n \cdot ||A_n(\tau)|^2-I^{GS}(\tau)|}{\sum_n
||A_n(\tau)|^2-I^{GS}(\tau)|}\;,V(\tau)=\dot{X}(\tau)\;.
\end{equation}
Now we note, that any single-harmonic variation of $\phi_{in}(\tau)=\phi_M \sin [\omega (\tau-\tau_0)]$
possesses the time-shift symmetry
\begin{equation}
\phi_{in}(\tau+T/2)=-\phi_{in}(\tau)\;.
\label{timeshift}
\end{equation}
Provided this symmetry holds, the dynamical
equation (\ref{DNLS}) remains invariant under the operation $\hat{S}_a$ 
\begin{equation}
\hat{S}_a:\; \tau \to \tau+T/2,\; n\to -n .
\label{symmetry}
\end{equation}
%
At the
same time, this operation changes the sign of the soliton
velocity. Therefore, no rectification of a single-harmonic AC force is
possible with cavity solitons. However, the above symmetry can be broken e.g. by
choosing a biharmonic variation of the incidence angle:
\begin{equation}
\label{biharm1}
\phi_{in}(\tau)=
\phi_a \sin (\omega \tau) + \phi_b \sin (2\omega \tau + \theta)\;.
\end{equation}
Yet another possibility is to use two holding beams, which oscillate at the first
and second harmonic frequency, respectively:
\begin{eqnarray}
\label{biharm2}
A^{in}_n=a\exp\left[i\phi_{in}^{(1)}(\tau)n\right]+b\exp\left[i\phi_{in}^{(2)}(\tau)n\right]\;,\\
\label{biharm_phase}
\phi_{in}^{(1)}=\phi_{a}  \sin (\omega \tau)n\;,\quad
\phi_{in}^{(2)}=\phi_{b}  \sin (2\omega \tau+\theta)n\;.
\end{eqnarray}
In both cases a unidirectional propagation of DCSs 
can be observed, despite the fact that the averaged value of the incidence
angle (and thus that of the force acting on the DCS) is zero, see
Fig.~\ref{fig3}.
The resulting average velocity of the soliton strongly depends on the system parameters and parameters
of shaking beam(s) [cf. maximum velocities of dark and bright DCSs for different holding 
beams and coupling parameters in Fig.~\ref{fig4}]. 
One of the possible ways to control the DCS motion is to adjust the relative phase 
$\theta$ in (\ref{biharm1}),(\ref{biharm2}), 
see Fig.~\ref{fig4}. 

For slowly shaking holding beams the ratchet effect can be estimated through the DCS velocities as functions of
constant tilt $\mathcal{V}(\phi_{in})$.
%
%
The resulting average velocity of a DCS can be
obtained as
\begin{equation}
V=\frac1T\int_{0}^{T} \mathcal{V}[\phi_{in}(\tau)]d\tau
\label{vel_aver}
\end{equation}
We note, that in the continuum limit of Eq.~(\ref{DNLS}) the soliton velocity $\mathcal{V}(\phi_{in})$ 
is a linear function of the incidence angle \cite{Fedorov}. Thus, for any choice of periodic functions $\phi_{in}$ with zero mean 
the resulting average velocity (\ref{vel_aver}) is zero in the adiabatic limit. The ratchet effect for slowly varying
incidence angle(s) $\phi_{in}$ can appear only for nonlinear functions $\mathcal{V}(\phi_{in})$. 
This nonlinearity is induced by the discreteness of the system due to Peierls-Nabarro potential \cite{Pesh_mov}.
In that case the resulting soliton net motion does not depend on the actual choice of the frequency $\omega$ which
can be taken arbitrary small.


To conclude, we have demonstrated the possibility to generate a rectified motion 
of discrete cavity solitons by means of 
shaking holding beams with zero average inclination. 
The necessary condition for the observation of  
this soliton ratchet effect is the violation of the symmetry $\hat{S}_a$~(\ref{symmetry})
of the underlying model equations. This can be done e.g. by bi-harmonic variation
of the holding beam inclination angle~(\ref{biharm1}) or by a superposition of two
shaking holding beams with different frequencies. The velocity of the resulting soliton net motion
can be adjusted by the parameters of the holding beams ($\omega, \phi_a, \phi_b, \theta$).
This opens a new prominent perspective for soliton steering and various all-optical switching schemes.
Our results are also instructive for the general problem of nontopological
soliton ratchets in spatially extended discrete 
and continuous systems. 


\end{document}